\documentclass{Interspeech2024}

\interspeechcameraready

\usepackage{microtype}
\usepackage{xcolor}
\usepackage{pifont}

\usepackage{booktabs}
\usepackage{multirow}
\usepackage{tabularx}
\newcommand{\mytable}{
	\centering
	\renewcommand{\arraystretch}{1.2}
}
\newcolumntype{C}{>{\centering\arraybackslash}X}
\newcolumntype{L}{>{\raggedright\arraybackslash}X}
\newcolumntype{R}{>{\raggedleft\arraybackslash}X}
\newcolumntype{P}[1]{>{\raggedright\arraybackslash}p{#1}}
\usepackage{siunitx}
\usepackage{etoolbox}                           
\newcommand{\ubold}{\fontseries{b}\selectfont}  
\robustify\ubold                                

\graphicspath{{fig/}}

\usepackage{amsmath}
\DeclareMathOperator*{\argmin}{arg\,min}

\usepackage{cite}
\let\oldbibliography\thebibliography
\renewcommand{\thebibliography}[1]{
    \oldbibliography{#1}
    \setlength{\itemsep}{0.4pt}
    \vspace*{-0mm}
}

\title{Revisiting speech segmentation and lexicon learning with better features}
\name{Herman}{Kamper}
\name{Benjamin}{van Niekerk}
\address{
    Electrical and Electronic Engineering, Stellenbosch University, South Africa
}
\email{kamperh@sun.ac.za, benjamin.l.van.niekerk@gmail.com}
\keywords{
    unsupervised word segmentation, lexicon learning, zero-resource speech processing, self-supervised learning
}

\usepackage[prependcaption,textsize=scriptsize]{todonotes}
\definecolor{mycolor}{HTML}{FF6600}

\begin{document}

\maketitle
 
\begin{abstract}
    We revisit a self-supervised method that segments unlabelled speech into word-like segments. We start from the two-stage duration-penalised dynamic programming method that performs zero-resource segmentation without learning an explicit lexicon. In the first acoustic unit discovery stage, we replace contrastive predictive coding features with HuBERT. After word segmentation in the second stage, we get an acoustic word embedding for each segment by averaging HuBERT features. These embeddings are clustered using K-means to get a lexicon. The result is good full-coverage segmentation with a lexicon that achieves state-of-the-art performance on the ZeroSpeech benchmarks.
\end{abstract}

\section{Introduction}

Learning to segment and cluster unlabelled speech into word-like units could be valuable for developing new zero-resource speech technology and for cognitive models of language acquisition~\cite{dupoux_cognition18}.
Early methods did segmentation while clustering speech into an inventory of words~\cite{kamper+etal_asru17}.
But more recent self-supervised methods do segmentation without explicitly learning a lexicon~\cite{kamper2023word,algayres2022dp}.

\begin{figure}[!b]
    \centering
    \includegraphics[width=0.9\linewidth]{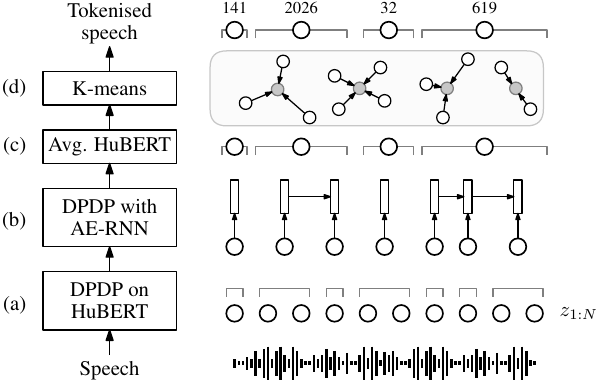}
    \caption{
        Starting from~\cite{kamper2023word}, we replace CPC with a HuBERT clustering model for acoustic unit discovery (a).
        After word segmentation (b), we construct a lexicon through K-means clustering (d) on averaged HuBERT acoustic word embeddings~(c).
    }
    \label{fig:method}
\end{figure}

In this paper we revisit the model from~\cite{kamper2023word} and propose two extensions: (1)~we incorporate better self-supervised features to do segmentation, and (2)~we add a clustering step to build a lexicon from the discovered segments.
The original approach in~\cite{kamper2023word} is a two-stage method that first discovers acoustic units by grouping speech into phone-like segments using duration-penalised dynamic programming (DPDP).
In the second stage, these acoustic units are divided into word-like segments using an autoencoding recurrent neural network (AE-RNN) as a scoring function with DPDP.
The two stages are illustrated at (a) and (b) in Fig.~\ref{fig:method}.
While the first stage learns a dictionary of phone-like units, the second stage in the original model performs segmentation directly without relying on discrete word categories.

Our changes are also illustrated in Fig.~\ref{fig:method}.
In the first stage~(a), we replace contrastive predictive coding (CPC)~\cite{vandenoord+etal_arxiv18} with HuBERT~\cite{hsu2021hubert} input features.
The latter performs better in many speech tasks~\cite{mohamed2022self}.
In the second stage~(b), segmentation is then performed on the quantised HuBERT representations.
To build a lexicon, we extract acoustic word embeddings by averaging HuBERT features over each word segment (c). 
Then we cluster the embeddings using K-means (d).

We apply this approach to the five languages in Track 2 of the ZeroSpeech Challenge: English, French, Mandarin, German and Wolof~\cite{dunbar+etal_interspeech20}.
Compared to existing full-coverage systems, our revised model gives the best lexicon and segmentation results.

\begin{table*}[!t]
    \caption{Performance (\%) for word segmentation (boundary and token $F_1$) and lexicon learning (normalised edit distance, NED) for full-coverage systems that also learn a lexicon on Track 2 of the ZeroSpeech Challenge.}
    \vspace*{-5pt}
    \label{tbl:results}
    \mytable
    \eightpt
    \begin{tabularx}{\linewidth}{@{}lCCCCCCCCCCCCCCCC@{}}
        \toprule
        & \multicolumn{3}{c}{English} & \multicolumn{3}{c}{French} & \multicolumn{3}{c}{Mandarin} & \multicolumn{3}{c}{German} & \multicolumn{3}{c}{Wolof}  \\
        \cmidrule(l){2-4} \cmidrule(l){5-7} \cmidrule(l){8-10} \cmidrule(l){11-13} \cmidrule(l){14-16}
        Model & NED & Bound.\ $F_1$ & Token $F_1$ & NED & Bound.\ $F_1$ & Token $F_1$ & NED & Bound.\ $F_1$ & Token $F_1$ & NED & Bound.\ $F_1$ & Token $F_1$ & NED & Bound.\ $F_1$ & Token $F_1$  \\
        \midrule
        Self-expressing AE~\cite{bhati+etal_interspeech20}
        & 89.5 & 40.1 & \hphantom{0}6.6 & 89.0 & 41.9 & \hphantom{0}6.3 & 96.6 & 52.2 & 12.1 & 89.6 & 36.2 & \hphantom{0}6.3 & 82.7 & 50.6 & 12.6 \\
        ES-KMeans~\cite{kamper+etal_asru17}
        & 73.2 & 56.7 & 19.2 & 68.7 & 43.3 & \hphantom{0}6.3 & 88.1 & 54.5 & \hphantom{0}8.1 & 66.2 & 45.2 & 11.5 & 72.4 & 52.8 & 10.9\\
        DPDP HuBERT [ours]
        & \textbf{41.7} & \textbf{58.0} & \textbf{19.6} & \textbf{66.0} & \textbf{53.2} & \textbf{11.6} & \textbf{86.0} & \textbf{67.1} & \textbf{24.5} & \textbf{56.8} & \textbf{50.3} & \textbf{12.5} & \textbf{72.2} & \textbf{61.2} & \textbf{13.1} \\
        \addlinespace
        VG-HuBERT~\cite{peng+harwath_interspeech22} & 
        41.0 & 59.0 & 24.0 & 62.0 & 55.0 & 15.0 & 73.0 & 62.0 & 19.0 & 56.0 & 50.0 & 15.0 & 92.0 & 52.0 & \hphantom{0}9.0 
        \\
        \bottomrule
    \end{tabularx}
\end{table*}

\section{Method}

We now describe the method from Fig.~\ref{fig:method} in more detail.
For acoustic unit discovery (a), we use constrained clustering on a sequence of HuBERT features $\mathbf{x}_{1:T}$.
The sequence is broken up into segments, with all frames in a potential segment $\mathbf{x}_{a:b}$ assigned to a cluster centroid from a codebook of acoustic units $\left\{ \mathbf{e}_k \right\}_{k = 1}^K$.
Formally, $\mathbf{x}_{1:T}$ is divided into contiguous segments $s_{1:N}$, with each segment $s_n = (a_n, b_n, z_n)$ specified by a start frame $a_n$, and end frame $b_n$, and the code index $z_n$.
We then use dynamic programming to find the segmentation that minimises the combined sum of squared distances:
\vspace*{-5pt}
\begin{equation}
    \argmin_{s_{1:N}} \sum_{s_n \in s_{1:N}}
        \sum_{t = a_n}^{b_n} \lVert \mathbf{x}_t  - \mathbf{e}_{z_t} \rVert^2
        - \lambda (b_n - a_n)
    \label{eq:dpdp}
    \vspace*{-2pt}
\end{equation}
The second term is a duration penalty encouraging longer segments with $\lambda$ controlling its strength. 
The idea is to partition the features by minimising the distance between the frames within a segment.
The result is a sequence of code indices $z_{1:N}$ corresponding to hypothesised acoustic units (bottom-most circles in Fig.~\ref{fig:method}).
For this stage we use features from the seventh layer of HuBERT-Base since it correlated best with phone labels in~\cite{pasad2023comparative}.

Next, we segment the sequence of code indices $z_{1:N}$ into word-like units (b).
This is again done through DPDP, but now instead of using a clustering model to score segments, an AE-RNN is used.
This encoder-decoder model takes a sequence of code indices, summarises it into a single embedding vector with an RNN, conditions a decoder RNN with the embedding, and tries to reconstruct the input sequence.
Some input segments will be reconstructed well (with the idea that these would correspond to words) while others will not (non-words). So we can use the AE-RNN as the DPDP scoring function.
Formally, the inner sum in~\eqref{eq:dpdp} is replaced with $-\sum_{t = a_n}^{b_n} \log P_{\textrm{RNN}}(z_t|z_{a_n:b_n})$ and we again use a duration penalty to encourage longer word segments.
In practice, the AE-RNN is trained on complete full-length utterances (encoded beforehand into code sequences) and then fixed for subsequent DPDP segmentation.

The result up to now are predicted word boundaries: the grey bars at (c) in Fig.~\ref{fig:method}.
But a lexicon is not learned.
We extend~\cite{kamper2023word} to address this.
First we just tried to K-means cluster the latent AE-RNN embeddings, but this did not perform competitively on Buckeye development data.
Based on~\cite{sanabria2023analyzing}, we settled on simply getting acoustic word embeddings by averaging the HuBERT features in each hypothesised segment (c) and then K-means clustering these (d).
For this we use the ninth HuBERT-Base layer since this correlated best with word labels in~\cite{pasad2023comparative}.

\section{Experimental setup}

\hspace{\the\parindent} \textbf{Data and metrics.}
We run experiments on Track 2 of the ZeroSpeech Challenge~\cite{dunbar+etal_interspeech20}.
The challenge covers five languages: English, French, Mandarin, German and Wolof with 45, 24, 2.5, 25 and 10 hours of speech, respectively.
Segmentation is evaluated with boundary $F_1$ and token $F_1$ scores; the latter requires both boundaries of a word to be correct.
To assess the lexicon, each inferred word token is mapped to its overlapping phoneme sequence using aligned transcriptions. 
Then we calculate the normalised edit distance (NED) between phoneme sequences for all pairs of segments in the same cluster (lower is better).

\textbf{Our method.}
We use the English HuBERT-Base model for all languages.
We have four hyperparameters: the acoustic unit codebook size, the lexicon size, and the duration penalty $\lambda$ for each of the two stages. We use 100 acoustic units and didn't try any other setting.
For a fair comparison, we select the lexicon size to match ES-KMeans~\cite{kamper+etal_asru17}: 43k, 29k, 3k, 29k and 3.5k for English, French, Mandarin, German and Wolof, respectively.
ES-KMeans itself selected its lexicon size as a fraction of the number of automatically detected syllable tokens in each language.
Based on development experiments on Buckeye, we set $\lambda = 2$ for acoustic unit discovery and $\lambda = 5$ for word segmentation for all languages.

\textbf{Baselines.}
We compare our method to three other submissions to Track 2 of the ZeroSpeech Challenge. 
We restrict the comparison to full-coverage systems that also build an explicit lexicon.
ES-KMeans combines segmentation with K-means clustering on averaged MFFCs~\cite{kamper+etal_asru17}.
The self-expressing autoencoder uses a graph-growing clustering algorithm to build its lexicon on learned features~\cite{bhati+etal_interspeech20}.
We also report scores for the VG-HuBERT method~\cite{peng+harwath_interspeech22}, but because this approach doesn't exclusively use unlabelled audio (it uses images paired with speech), we do not consider this as a direct comparison.

\section{Experiments}

Results are shown in Table~\ref{tbl:results}.
Of the full-coverage systems that make exclusive use of unlabelled speech, our new approach gives competitive segmentation scores ($F_1$) and produces the best lexicons (NED) across all five languages.
In some cases we even outperform the visually grounded approach~\cite{peng+harwath_interspeech22}.

The original CPC-based DPDP method~\cite{kamper2023word} doesn't build an explicit lexicon, but we can still look at the segmentation scores:
these aren't given in the table, but they are very similar to DPDP HuBERT.
It therefore seems that the better HuBERT features are mostly beneficial for word category learning.
(We could construct a model with CPC in one part and HuBERT in another, but it is preferable to have a model with one feature type.)

\section{Conclusion}

We replaced CPC features with better HuBERT features in an existing model for segmenting speech and then added a clustering model on top to learn a lexicon.
This gave some of the best results on the ZeroSpeech Challenge benchmarks.

\bibliography{dpdp_lexicon_2024}

\end{document}